\documentclass[preprint,pra,showpacs,showkeys]{revtex4}
\usepackage{graphicx,amsmath,amsfonts,amssymb}
\usepackage{times}
\begin{document}
\title{Atomic entanglement sudden death in a strongly driven cavity QED system}
\author{Ying-Jie Zhang, Zhong-Xiao Man}
\author{Yun-Jie Xia \footnote{ The corresponding author. Email address:
yjxia@mail.qfnu.edu.cn}} \affiliation {Shandong Provincial Key
Laboratory of Laser Polarization and Information Technology,
Department of Physics, Qufu Normal University, Qufu 273165, Peoples
Republic of China }

\begin{abstract}
We study the entanglement dynamics of strongly driven atoms
off-resonantly coupled with cavity fields. We consider conditions
characterized not only by the atom-field coupling but also by the
atom-field detuning. By studying two different models within the
framework of cavity QED, we show that the so-called atomic
entanglement sudden death (ESD) always occurs if the atom-field
coupling larger than the atom-field detuning, and is independent of
the type of initial atomic state.
\end{abstract}

\maketitle

\section*{$1.$ \textbf{Introduction}}
Entanglement plays a crucial role in quantum information processing
$[1]$. Quantum algorithms (particularly, in Shor's algorithm, to
find the prime factors of an $n$-bit integer) exploit entanglement
to speed up computation. Entanglement dynamics have been a difficult
subject and have attracted extensive interest recently ranging from
two-qubit systems $[2-5]$, to continuous variables $[6,7]$, spin
systems $[8,9,10]$, and multi-partite systems $[11-14]$.
Entanglement of open quantum systems also has attracted considerable
attention due to its significance for both fundamentals and
applications of quantum information processing $[15,16]$. Such as,
the paper $[15]$ gives us the non-Markovian entanglement evolution
of two uncoupled qubits which are coupled to their own independent
environments. Origination and survival of entanglement for two
independent qudits dephasingly coupled to a common zero-temperature,
super-Ohmic, bosonic environment is studied in $[16]$. Moreover, in
the close systems, proposals $[17,18]$ have been made for the direct
measurement of finite-time disentanglement in cavity QED, and
real-time detection of entanglement sudden death (ESD) has been
reported very
recently$[19,20,21]$. \\
\indent The interaction of a two-level atom with a quantized single
mode of a harmonic oscillator, was called the Jaynes-Cummings (JC)
model $[22]$. The JC model has found its natural playground in the
field of cavity quantum electrodynamics (CQED), and extensions of
the JC model to more atoms and more modes, externally driven or not,
have been developed. Presently, we enjoy a vast number of
theoretical and experimental developments. In $[19]$, the authors
dealt with a double JC model in which two initially entangled
two-level atoms $A$ and $B$ are independently coupled with separate
cavity fields $a$ and $b$ respectively, but there are no
interactions at all between the subsystems $Aa$ and $Bb$. Focusing
on the atomic subsystem they found that depending on the type of
initial state of atoms $AB$, their entanglement may or may not
exhibit ESD. From $[19]$, we can acquire that the authors have taken
Bell-like states as the initial state of atoms and assumed the
initial cavity field to be in vacuum. As a consequence, ESD was
found to be sensitive to the initial atomic state. That is to say,
ESD may occur for a certain type of initial
atomic state but does not appear for another type. \\
\indent In this paper, we consider a system consisting of a
two-level atom trapped inside a single mode cavity. The atom is
additionally driven by a strongly classical field. The experimental
implementation seems to be feasible due to the recent advances in
deterministic trapping of atoms in the optical cavities $[23,24]$.
We show that under-taking atomic pure Bell-like states
$|\Phi_{AB}\rangle=\cos\theta|e_{A},e_{B}\rangle+\sin{\theta}e^{i\phi}|g_{A},g_{B}\rangle$
or
$|\Psi_{AB}\rangle=\cos\theta|e_{A},g_{B}\rangle+\sin{\theta}e^{i\phi}|g_{A},e_{B}\rangle$
(with $0\leq\theta\leq2\pi$ and $0\leq\phi\leq\pi$) as the initial
state of atoms and assuming the initial cavity field to be in
vacuum, ESD can occur for two types of initial atomic states. In
order to study entanglement we will use the $negativity$ ($N$),
which can be defined for two qubits as two times the modulus of the
negative eigenvalue of the partial transposition of the state
$\rho$, $\rho^{T_{A}}$ $[25,26]$, if it exists. For short
\begin{equation}
N(\rho)=2\max\{{0,-\lambda_{min}}\},\label{1}
\end{equation}
where $\lambda_{min}$ is the lowest eigenvalue of $\rho^{T_{A}}$.
Our choice is motivated by the facts that the negativity is easy to
calculate and provides full entanglement information for a two-qubit
system. We also need to quantify three-party entanglement in the
tree-qubit state $|\psi_{ABa}\rangle$, the three-tangle
$\tau_{3}(\psi_{ABa})$ has been introduced in Ref.$[27]$. It can be
expressed by using the wave function coefficients
$\{\psi_{000},\psi_{001},\cdots,\psi_{111}\}$ as
\begin{eqnarray}
\tau_{3}&=&4|d_{1}-2d_{2}+4d_{3}|,\nonumber\\
d_{1}&=&\psi_{000}^{2}\psi_{111}^{2}+\psi_{001}^{2}\psi_{110}^{2}+\psi_{010}^{2}\psi_{101}^{2}+\psi_{100}^{2}\psi_{011}^{2},\nonumber\\
d_{2}&=&\psi_{000}\psi_{111}\psi_{011}\psi_{100}+\psi_{000}\psi_{111}\psi_{101}\psi_{010}+\psi_{000}\psi_{111}\psi_{110}\psi_{001}\nonumber\\
&+&\psi_{011}\psi_{100}\psi_{101}\psi_{010}+\psi_{011}\psi_{100}\psi_{110}\psi_{001}+\psi_{101}\psi_{010}\psi_{110}\psi_{001},\nonumber\\
d_{3}&=&\psi_{000}\psi_{110}\psi_{101}\psi_{011}+\psi_{111}\psi_{001}\psi_{010}\psi_{100}.\label{01}
\end{eqnarray}
\indent To figure out general conditions for the possible negativity
of atomic ESD, we shall study two different atom-cavity models,
which we refer to as Model $1$ and Model $2$. Model $1$ is
considered in section $2$, in this model one of two atoms is trapped
in a single cavity, off-resonantly coupled to this cavity, and
driven by a classic strong coherent field, while the other remains
outside the cavity and has no environment. In section $3$ we
consider Model $2$ which deals with a double driven JCM: each of the
two strongly driven atoms interacts with its own cavity in the
absence of any coupling between the atom-field subsystems. A generic
result we have found out is that the atomic ESD always occurs in the
conditions of the atom-field coupling larger than the atom-field
detuning. Finally, we conclude in section $4$.
\section*{$2.$ \textbf{Model $1$}}
\indent In this section we show a model in which a strongly driven
atom $A$ is off-resonantly coupled to a single-mode cavity field
$a$, while the other atom $B$ is isolated from all environment. The
Hamiltonian of the system can be described by
\begin{equation}
H=\frac{\hbar{\omega}_{a}}{2}\sigma^{z}_{A}+\frac{\hbar{\omega}_{b}}{2}\sigma^{z}_{B}+\hbar\omega_{f}a^{\dag}a+\hbar\Omega(e^{-i\omega_{D}t}\sigma^{+}_{A}+e^{i\omega_{D}t}\sigma_{A})
+{\hbar}g(\sigma^{+}_{A}a+{\sigma}_{A}a^{\dag}),\label{2}
\end{equation}
where $\Omega$ is the Rabi frequency associated with the coherent
driving field amplitude, $g$ is the atom-cavity mode coupling
constant, $a(a^{\dag})$ the field annihilation (creation) operator,
$\sigma_{A}=|g_{A}\rangle{\langle}e_{A}|$
($\sigma^{+}_{A}=|e_{A}\rangle{\langle}g_{A}|$) the atomic lowering
(raising) operator, and
$\sigma^{z}_{A}=|e_{A}\rangle{\langle}e_{A}|-|g_{A}\rangle{\langle}g_{A}|$
the inversion operator. Considering the strong-driving regime for
the interaction between the atom and the external coherent field
$\Omega{\gg}\{g,\delta\}$ and making $\omega_{a}=\omega_{D}$, we can
use the rotating-wave approximation (RWA) obtaining the effective
Hamiltonian $[28]$
\begin{equation}
H_{eff}=\frac{{\hbar}g}{2}(\sigma^{+}_{A}+\sigma_{A})(ae^{i{\delta}t}+a^{\dag}e^{-i{\delta}t}).\label{3}
\end{equation}
where $\delta=\omega_{f}-{\omega}_{a}$  is the atom-cavity
detuning.\\
\indent We first choose the state $|\Psi_{AB}\rangle$ as an initial
state of atoms $A$ and $B$, and assume the initial cavity field to
be in vacuum ($|Vacuum\rangle$ means no photon in the single-mode
cavity field). Thus the total system state at $t=0$ is
\begin{equation}
|\Psi(0)\rangle_{ABa}=(\cos\theta|e_{A},g_{B}\rangle+\sin{\theta}e^{i\phi}|g_{A},e_{B}\rangle)\otimes|Vacuum\rangle_{a}.\label{4}
\end{equation}
For simplicity, we make $\phi=0$. Then the evolved state in time $t$
will be $|\Psi(t)\rangle=U(t)|\Psi(0)\rangle$, and
$U(t)=\exp(-iH_{eff}t)$, so
\begin{eqnarray}
|\Psi(t)\rangle_{ABa}&=&\frac{\cos\theta+\sin{\theta}}{2}|\alpha\rangle|+_{A},+_{B}\rangle+\frac{\cos\theta-\sin{\theta}}{2}|\alpha\rangle|+_{A},-_{B}\rangle\nonumber\\
&-&\frac{\cos\theta-\sin{\theta}}{2}|-\alpha\rangle|-_{A},+_{B}\rangle-\frac{\cos\theta+\sin{\theta}}{2}|-\alpha\rangle|-_{A},-_{B}\rangle,\label{5}
\end{eqnarray}
with $\alpha=\frac{g}{2\delta}(1-e^{i{\delta}t})$ and
$|\pm_{X}\rangle=(|g_{X}\rangle\pm|e_{X}\rangle)/\sqrt{2}$, where
$X=A,B$ and
$\{|i\rangle\}^{4}_{i=1}=\{|+_{A}+_{B}\rangle,|+_{A}-_{B}\rangle,|-_{A}+_{B}\rangle,|-_{A}-_{B}\rangle\}$
is the rotated basis of the atomic Hilbert space. We now try to
estimate the atomic entanglement of the state Equation(\ref{5}), and
we notice that the cavity field in the general nonorthogonal
coherent state $|\alpha\rangle$ and $|-\alpha\rangle$. So we can
define $|0\rangle=|\alpha\rangle$,
$|1\rangle=(|-\alpha\rangle-P|\alpha\rangle)/\sqrt{1-P^{2}}$ with
$P=\exp(\frac{g^{2}}{\delta^{2}}(\cos{\delta}t-1))$ for the cavity
field subsystem. Then Equation(\ref{5}) can be written as
\begin{eqnarray}
|\Psi(t)\rangle&=&(\frac{\cos\theta+\sin{\theta}}{2}|+_{A},+_{B}\rangle+\frac{\cos\theta-\sin{\theta}}{2}|+_{A},-_{B}\rangle\nonumber\\
&-&\frac{\cos\theta-\sin{\theta}}{2}P|-_{A}+_{B}\rangle-\frac{\cos\theta+\sin{\theta}}{2}P|-_{A}-_{B}\rangle)|0\rangle\nonumber\\
&-&(\frac{\cos\theta-\sin{\theta}}{2}\sqrt{1-P^{2}}|-_{A}+_{B}\rangle+\frac{\cos\theta+\sin{\theta}}{2}\sqrt{1-P^{2}}|-_{A}-_{B}\rangle)|1\rangle\label{6}
\end{eqnarray}
In order to calculate the negativity of atoms, we trace out the mode
$a$, the expression for $\rho^{\Psi}_{AB}(t)$ in the rotated basis
of the atomic Hilbert space
\begin{eqnarray}
\rho_{11}&=&\rho_{44}=(\frac{\cos\theta+\sin\theta}{2})^{2},\rho_{22}=\rho_{33}=(\frac{\cos\theta-\sin\theta}{2})^{2},\nonumber\\
\rho_{14}&=&\rho_{41}=-(\frac{\cos\theta+\sin\theta}{2})^{2}P,\rho_{23}=\rho_{32}=-(\frac{\cos\theta-\sin\theta}{2})^{2}P,\nonumber\\
\rho_{12}&=&\rho_{21}=\rho_{34}=\rho_{43}=\frac{\cos^{2}\theta-\sin^{2}\theta}{4},\nonumber\\
\rho_{13}&=&\rho_{31}=\rho_{42}=\rho_{24}=-\frac{\cos^{2}\theta-\sin^{2}\theta}{4}P.\label{7}
\end{eqnarray}
The negativity of $\rho^{\Psi}_{AB}(t)$ is
\begin{equation}
N(\rho^{\Psi}_{AB}(t))=2\max\{0,-\frac{1}{4}(1-P-\sqrt{1+P^{2}-2P\cos4\theta})\}.\label{8}
\end{equation}
\begin{figure}
\includegraphics[scale=1.2]{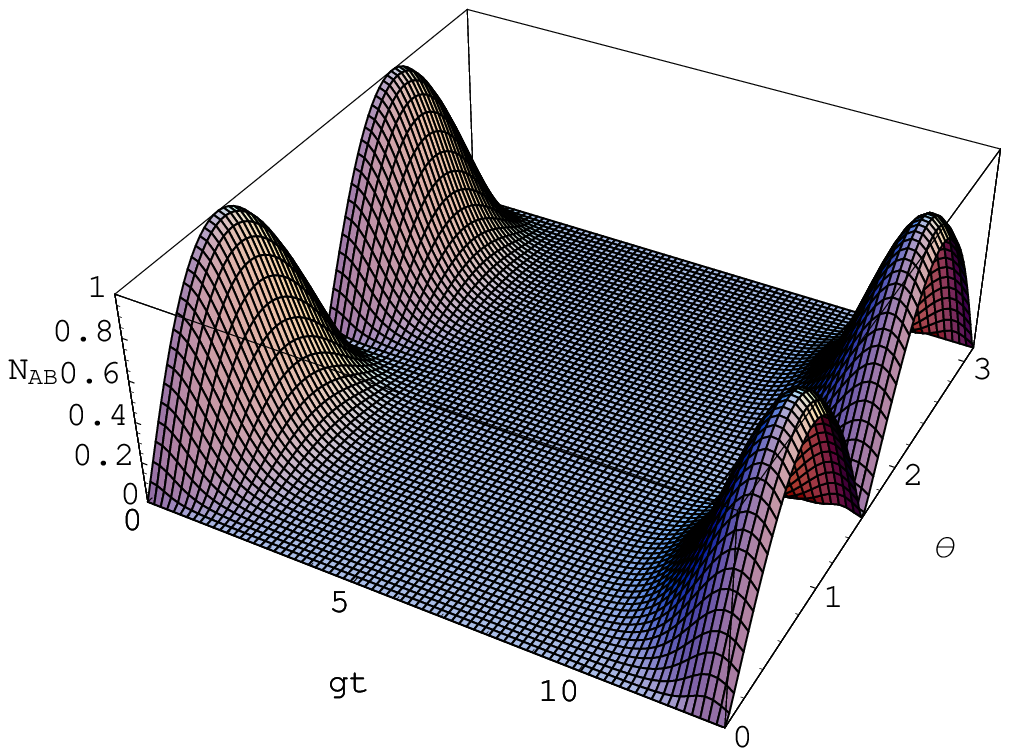}
\caption{\label{fig1}$N(\rho^{\Psi}_{AB}(t))$ as a function of
$\theta$ and the dimensionless time $gt$ for $g/\delta=2$, with
$\delta=0.5g$ in Model $1$.}
\end{figure}
\begin{figure}
\includegraphics[scale=1.2]{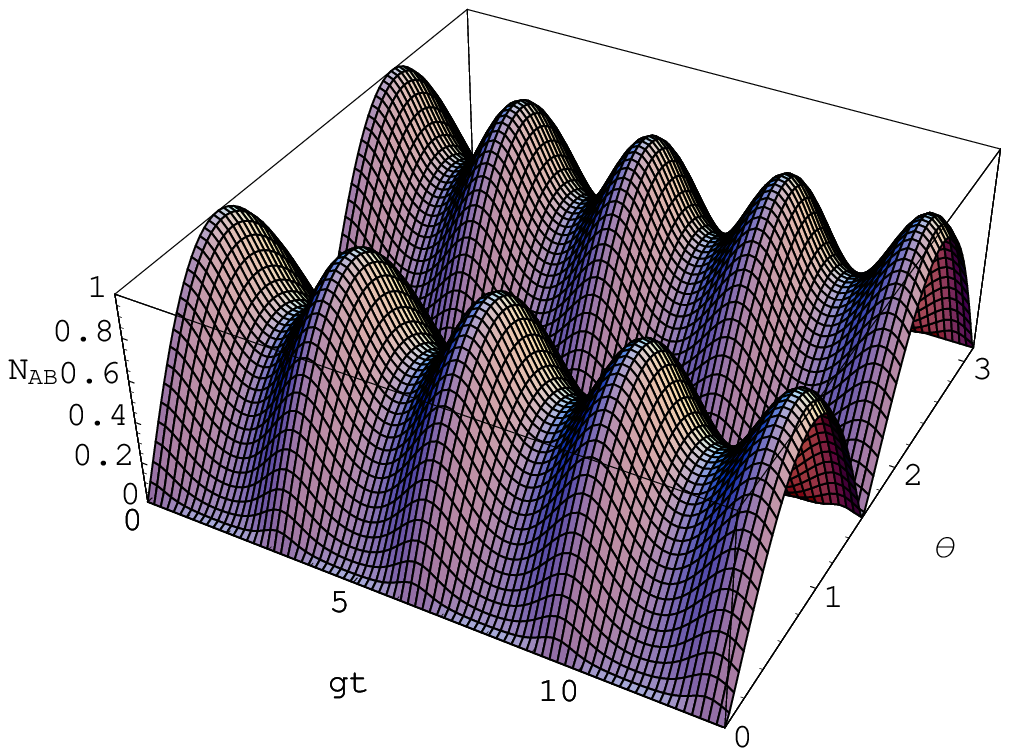}
\caption{\label{fig2}$N(\rho^{\Psi}_{AB}(t))$ as a function of
$\theta$ and the dimensionless time $gt$ for $g/\delta=0.5$, with
$\delta=2g$ in Model $1$.}
\end{figure}
\begin{figure}
\includegraphics[scale=1.2]{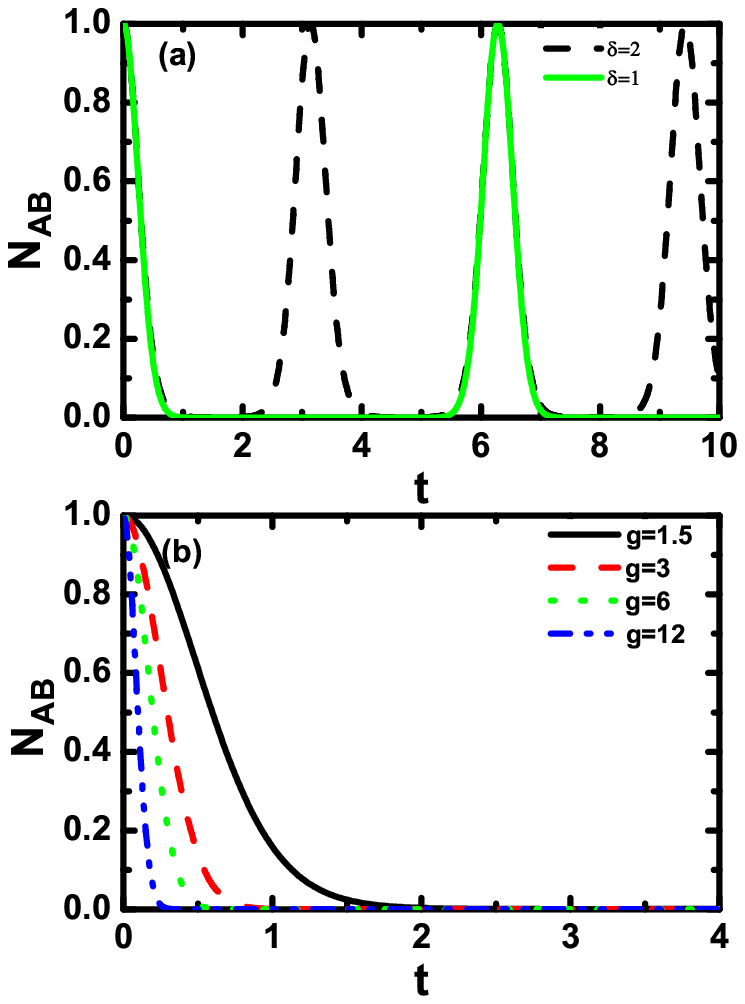}
\caption{\label{fig3} (a) $N(\rho^{\Psi}_{AB}(t))$ as a function of
time $t$ for $g=4$, $\theta=\pi/4$ in Model $1$, (b)
$N(\rho^{\Psi}_{AB}(t))$ as a function of time $t$ for $\delta=1$,
$\theta=\pi/4$ in Model $1$.}
\end{figure}
\indent The negativity $N(\rho^{\Psi}_{AB}(t))$ has been calculated,
as shown in Fig.$1$ and Fig.$2$, for the different $g/\delta$.
Obviously the negativity is fluctuating with dimensionless time $gt$
and the mixing angle $\theta$, and it can be zero in a finite amount
of time (the so-called entanglement sudden death). An important
point is that ESD is sensitive to the value of $g/\delta$ as
displayed in Fig.$1$ and Fig.$2$. For $g/\delta>1$, ESD happens
readily (Fig.$1$), and the larger the value of $g/\delta$ is, the
more easily the ESD appears. When $g/\delta\leq1$, ESD will not
appear at any time (Fig.$2$). This phenomenon shows that ESD is
related to $g/\delta$ completely in this model. Moreover, As we
study the entanglement dynamics under a close system, Fig.$1$ also
shows that sudden death and resurrection of atomic entanglement can
occur periodically as well as alternately. After entanglement of two
atoms occurs sudden death for a finite time, atomic entanglement can
revive again gradually. We have verified that when the ratio
$g/\delta$ is larger than the one used in Fig.$1$, the region of ESD
and entanglement revivals increases, and they last for a longer
time. On the other hand, when we decrease $g/\delta$, the ESD
regions shrink and the period of the entanglement oscillations becomes smaller.\\
\begin{figure}
\includegraphics[scale=1.2]{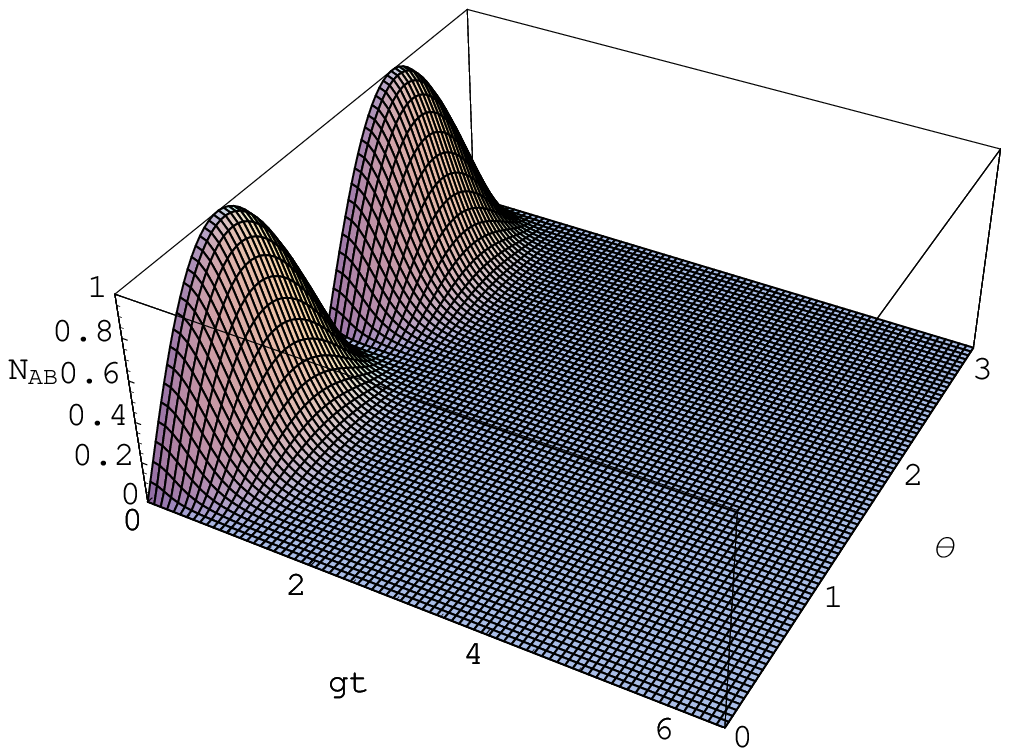}
\caption{\label{fig4} $N(\rho^{\Psi}_{AB}(t))$ as a function of
$\theta$ and the dimensionless time $gt$ for $\delta=0$ in Model
$1$}
\end{figure}
\begin{figure}
\includegraphics[scale=1.2]{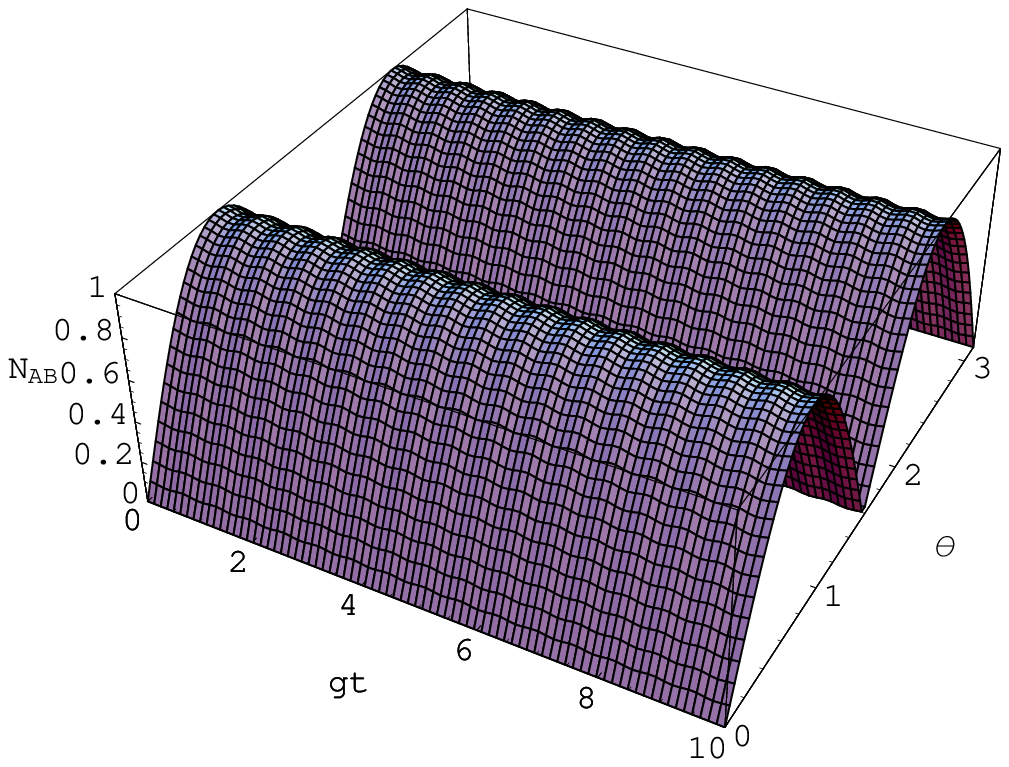}
\caption{\label{fig5} $N(\rho^{\Psi}_{AB}(t))$ as a function of
$\theta$ and the dimensionless time $gt$ for $\delta=10g$, in the
large detuning regime in Model $1$.}
\end{figure}
 \indent The character revealed by Fig.$3$ is that $\delta$
 influences the period of the ESD and $g$ is related to the velocity
of the $AB$ subsystem's disentanglement. If we want to acquire a
long-time ESD, then we can choose a smaller atom-field detuning
$\delta$(Fig.$3(a)$). While $\delta$ is a constant, the $AB$
subsystem can rapidly disentangle with the larger $g$ (Fig.$3(b)$).
A physical interpretation of the result is that the more the atom to
the cavity mode couples, the easier the initial entanglement decays.
Then we consider the atom and the cavity under resonance condition
$(\delta=0)$ or the large detuning regime $(\delta{\gg}g)$. Firstly,
in the resonant regime the negativity of $\rho^{\Psi}_{AB}(t)$
decreases at the beginning, and then vanishes at any time and
resurrection of atomic entanglement can not happen forever under
this condition(Fig.$4$), so the strongly driven model under the
resonant regime can be considered as an information eraser.
Secondly, entangled states are generally very fragile against
interaction with environments, but in our present scheme, the energy
exchange between atom and cavity do not exist under the atom-cavity
large detuning regime, so the initial entanglement of atoms can be
preserved during system evolution, which is shown as Fig.$5$. That
is to say, the strongly driven and large detuning model can be used
to preserve the information
(entanglement) in quantum processing and quantum computing.\\
\indent Now we move to the negativity $N(\rho^{\Psi}_{Aa}(t))$ and
$N(\rho^{\Psi}_{Ba}(t))$ and the three-tangle
$\tau(\rho^{\Psi}_{ABa}(t))$, and we acquire the following
\begin{equation}
N(\rho^{\Psi}_{Aa}(t))=2\max\{0,Q_{Aa}\},
\end{equation}
\begin{equation}
N(\rho^{\Psi}_{Ba}(t))=2\max\{0,Q_{Ba}\},
\end{equation}
\begin{equation}
\tau(\rho^{\Psi}_{ABa}(t))=\frac{1}{2}(1-P^{2})(1-\cos4\theta),
\end{equation}
\begin{figure}
\includegraphics[scale=1.6]{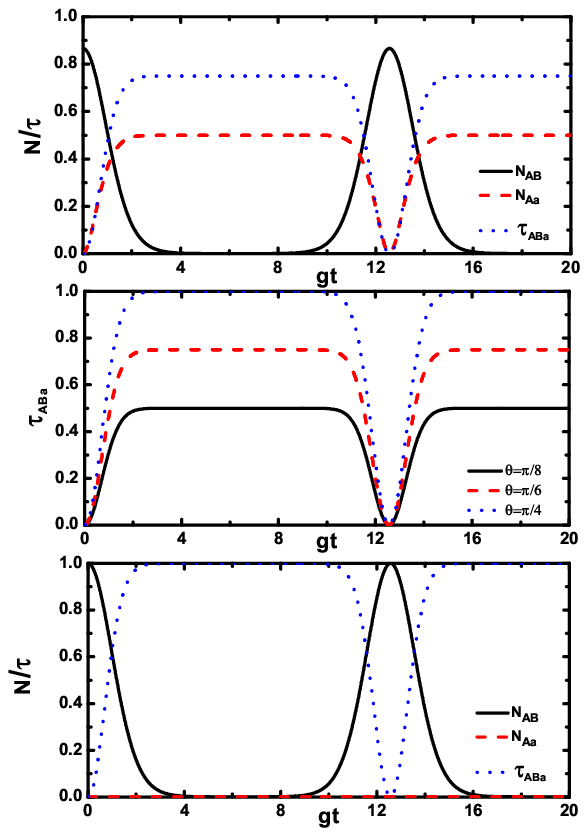}
\caption{\label{fig6}
$N(\rho^{\Psi}_{AB}(t))$/$N(\rho^{\Psi}_{Aa}(t))$/$\tau(\rho^{\Psi}_{ABa}(t))$
as a function of the dimensionless time $gt$ in Model $1$,
(a)$\delta=0.5g$, $\theta=\frac{\pi}{6}$, (b)$\delta=0.5g$, (c)
$\delta=0.5g$, $\theta=\frac{\pi}{4}$.}
\end{figure}
if $ 0<{\cos}2\theta<1$, then
$Q_{Aa}=-\frac{1}{8}(2-2{\cos}2\theta-\sqrt{2}\sqrt{3+(4-8P^{2}){\cos}2\theta+{\cos}4\theta})$
and $Q_{Ba}=-\frac{1}{2}(1-P{\cos}2\theta)$, and if
$-1<{\cos}2\theta<0$, then
$Q_{Aa}=-\frac{1}{8}(2+2{\cos}2\theta-\sqrt{2}\sqrt{3+(8P^{2}-4){\cos}2\theta+{\cos}4\theta})$
and $Q_{Ba}=-\frac{1}{2}(1+P{\cos}2\theta)$. Due to $0<P<1$, then
$Q_{Ba}<0$, the negativity $N(\rho^{\Psi}_{Ba}(t))=0$. In the
strongly driven regime, $B$ is isolated from all environment, and
the strong classical field that drives the atom $A$ can be used to
enhance the atom$(A)$-cavity field($a$) interaction. The
entanglement between $A$ and cavity $a$ can appear during the system
evolution, but $B$ and cavity $a$ can not entangle with each other
in the strongly driven regime(that is different from the standard
J-C Model $[21]$). It is interesting to find that while the $AB$
subsystem appears the long-time ESD, the entanglement of the $Aa$
subsystem and the three-tangle of the whole system are all on the
long-time invariable entanglement(as shown in Fig.$6(a)$). It is
clearly shown that the three-tangle $\tau_{ABa}$ can be influenced
by the initial state $|\Psi_{AB}(0)\rangle$ in Fig.$6(b)$. On the
condition $\theta=\pi/4$ (the initial state is on the maximum
entanglement), when $AB$ subsystem's entanglement vanishes for a
long time, the three-tangle of the whole system can achieve the
maximum value $1$ at the same time and the entanglement between $A$
and $a$ can not appear for any time. That is to say, the $AB$
subsystem's entanglement is transferred to the whole system's
entanglement thoroughly (Fig.$6(c)$).
So, we can acquire the three-partite system's long-time maximum entanglement in the strongly driven regime (In Model $1$).\\
 \indent Next, we choose $|\Phi_{AB}\rangle$ to be the
initial atomic state. Then the initial total system state is
\begin{equation}
\Phi_{ABa}(0)=(\cos\theta|e_{A},e_{B}\rangle+\sin{\theta}e^{i\phi}|g_{A},g_{B}\rangle)\otimes|Vacuum\rangle_{a},\label{9}
\end{equation}
with $\phi=0$, and for $\rho^{\Phi}_{ABa}(t)$ we have the
expressions as follows
\begin{eqnarray}
\rho_{11}&=&\rho_{44}=(\frac{\cos\theta+\sin\theta}{2})^{2},\rho_{22}=\rho_{33}=(\frac{\cos\theta-\sin\theta}{2})^{2},\nonumber\\
\rho_{14}&=&\rho_{41}=(\frac{\cos\theta+\sin\theta}{2})^{2}P,\rho_{23}=\rho_{32}=(\frac{\cos\theta-\sin\theta}{2})^{2}P,\nonumber\\
\rho_{12}&=&\rho_{21}=\rho_{34}=\rho_{43}=-\frac{\cos^{2}\theta-\sin^{2}\theta}{4},\nonumber\\
\rho_{13}&=&\rho_{31}=\rho_{42}=\rho_{24}=-\frac{\cos^{2}\theta-\sin^{2}\theta}{4}P.\label{10}
\end{eqnarray}
The negativity of $\rho^{\Phi}_{AB}(t)$ is satisfied with
\begin{equation}
N(\rho^{\Phi}_{AB}(t))=N(\rho^{\Psi}_{AB}(t)).\label{11}
\end{equation}
Equation(\ref{11}) reveals that the behavior of
$N(\rho^{\Phi}_{AB}(t))$ dependent on $g/\delta$ is equal to that of
$N(\rho^{\Psi}_{AB}(t))$. We also calculate the negativity
$N(\rho^{\Phi}_{Aa}(t))$ and $N(\rho^{\Phi}_{Ba}(t))$ and the
three-tangle $\tau(\rho^{\Phi}_{ABa}(t))$, and find that they are
all equal to $N(\rho^{\Psi}_{Aa}(t))$, $N(\rho^{\Psi}_{Ba}(t))$ and
$\tau(\rho^{\Psi}_{ABa}(t))$.\\
\indent Consequently, for Model $1$, where only one driven atom
interacts with its environment, if the atom is not driven and the
cavity field is in the vacuum initially, ESD does not occur either
for $\rho^{\Phi}_{AB}(0)$ or $\rho^{\Psi}_{AB}(0)$. In contrast,
however, if the atom is driven by a strong classical field and both
the atom-cavity coupling and the atom-cavity detuning satisfy
$g/\delta>1$, the atomic subsystem $AB$ always evolves via ESD,
independent of the type of initial atomic state(i.e. it never
matters if it is $\rho^{\Phi}_{AB}(0)$ or $\rho^{\Psi}_{AB}(0)$).
\section*{$3.$ \textbf{Model $2$}}
\indent The double JCM, which has been intensively investigated
recently $[19]$, has shown that if atoms $A$ and $B$ are prepared in
the $|\Phi_{AB}\rangle$-type Bell-like pure state initially, then
the atomic entanglement dies in a finite time and remains dead for
some time before reviving itself again, i.e., ESD occurs, whereas
the $|\Psi_{AB}\rangle$-type Bell-like pure state dose not exhibit
ESD at all. However, these investigations are confined to the
atom-field under full resonance conditions. Our purpose here is
focus on the differences that may appear on the conditions the atoms
$A$ and $B$ independently driven by a strongly external classical
field and the atom-cavity off resonance.\\
\indent In Model $2$, we consider two remote two-level atoms $A$ and
$B$ which are first prepared to be in an entangled state, and then
let each atom couple with a single-mode cavity respectively. During
the interaction with the single-mode cavity, the two atoms are
independently driven by a strongly classical field. In the
strong-driving regime, The effective Hamiltonian governing Model $2$
is of the form
\begin{equation}
H_{eff}=\sum_{k=A,B}H^{k}_{eff},\label{12}
\end{equation}
where
\begin{equation}
H^{k}_{eff}=\frac{{\hbar}g_{k}}{2}(\sigma^{+}_{k}+\sigma_{k})(a_{k}e^{i{\delta_{k}}t}+a^{\dag}_{k}e^{-i{\delta_{k}}t}).\label{13}
\end{equation}
The total system state at $t=0$ of the form
\begin{equation}
|\Psi(0)\rangle_{ABab}=(\cos\theta|e_{A},g_{B}\rangle+\sin{\theta}e^{i\phi}|g_{A},e_{B}\rangle)\otimes|Vacuum\rangle_{a}|Vacuum\rangle_{b}.\label{14}
\end{equation}
also with $\phi=0$, then the evolved state in time $t$ will be
\begin{eqnarray}
|\Psi(t)\rangle_{ABab}&=&\exp(-iH_{eff}t)|\Psi(0)\rangle_{ABab}\nonumber\\
&=&\frac{\cos\theta+\sin\theta}{2}|\alpha_{a},\beta_{b}\rangle|+_{A},+_{B}\rangle+\frac{\cos\theta-\sin\theta}{2}|\alpha_{a},-\beta_{b}\rangle|+_{A},-_{B}\rangle\nonumber\\
&-&\frac{\cos\theta-\sin\theta}{2}|-\alpha_{a},\beta_{b}\rangle|-_{A},+_{B}\rangle-\frac{\cos\theta+\sin\theta}{2}|-\alpha_{a},-\beta_{b}\rangle|-_{A},-_{B}\rangle,\label{15}
\end{eqnarray}
where
$\{|i\rangle\}^{4}_{i=1}=\{|+_{A}+_{B}\rangle,|+_{A}-_{B}\rangle,|-_{A}+_{B}\rangle,|-_{A}-_{B}\rangle\}$
is the rotated basis of the atomic Hilbert space. Similar to
Equation(\ref{5}), we also define
\begin{eqnarray}
|0_{a}\rangle&=&|\alpha_{a}\rangle,|1_{a}\rangle=(|-\alpha_{a}\rangle-P_{A}|\alpha_{a}\rangle)/\sqrt{1-P^{2}_{A}},\label{16}\\
|0_{b}\rangle&=&|-\beta_{b}\rangle,|1_{b}\rangle=(|\beta_{b}\rangle-P_{B}|-\beta_{b}\rangle)/\sqrt{1-P^{2}_{B}},\label{17}
\end{eqnarray}
where
$P_{A}=\exp(\frac{g^{2}_{A}}{\delta^{2}_{A}}(\cos{\delta_{A}}t-1))$,$P_{B}=\exp(\frac{g^{2}_{B}}{\delta^{2}_{B}}(\cos{\delta_{B}}t-1))$.
The reduced density matrix $\rho^{\Psi}_{AB}(t)$ is
\begin{eqnarray}
\rho_{11}&=&\rho_{44}=(\frac{\cos\theta+\sin\theta}{2})^{2},\rho_{22}=\rho_{33}=(\frac{\cos\theta-\sin\theta}{2})^{2},\nonumber\\
\rho_{14}&=&\rho_{41}=-(\frac{\cos\theta+\sin\theta}{2})^{2}P_{A}P_{B},\rho_{23}=\rho_{32}=-(\frac{\cos\theta-\sin\theta}{2})^{2}P_{A}P_{B},\nonumber\\
\rho_{12}&=&\rho_{21}=\rho_{34}=\rho_{43}=\frac{\cos^{2}\theta-\sin^{2}\theta}{4}P_{B},\nonumber\\
\rho_{13}&=&\rho_{31}=\rho_{42}=\rho_{24}=-\frac{\cos^{2}\theta-\sin^{2}\theta}{4}P_{A}.\label{18}
\end{eqnarray}
The negativity of $\rho^{\Psi}_{AB}(t)$ is
\begin{equation}
N(\rho^{\Psi}_{AB}(t))=2\max\{0,-\frac{1}{8}(2-2P_{A}P_{B}-\sqrt{2}\sqrt{(1+P^{2}_{A})(1+P^{2}_{B})+(P^{2}_{A}+P^{2}_{B}
-1-4P_{A}P_{B}-P^{2}_{A}P^{2}_{B})\cos4\theta})\}.\label{19}
\end{equation}
\begin{figure}
\includegraphics[scale=1.2]{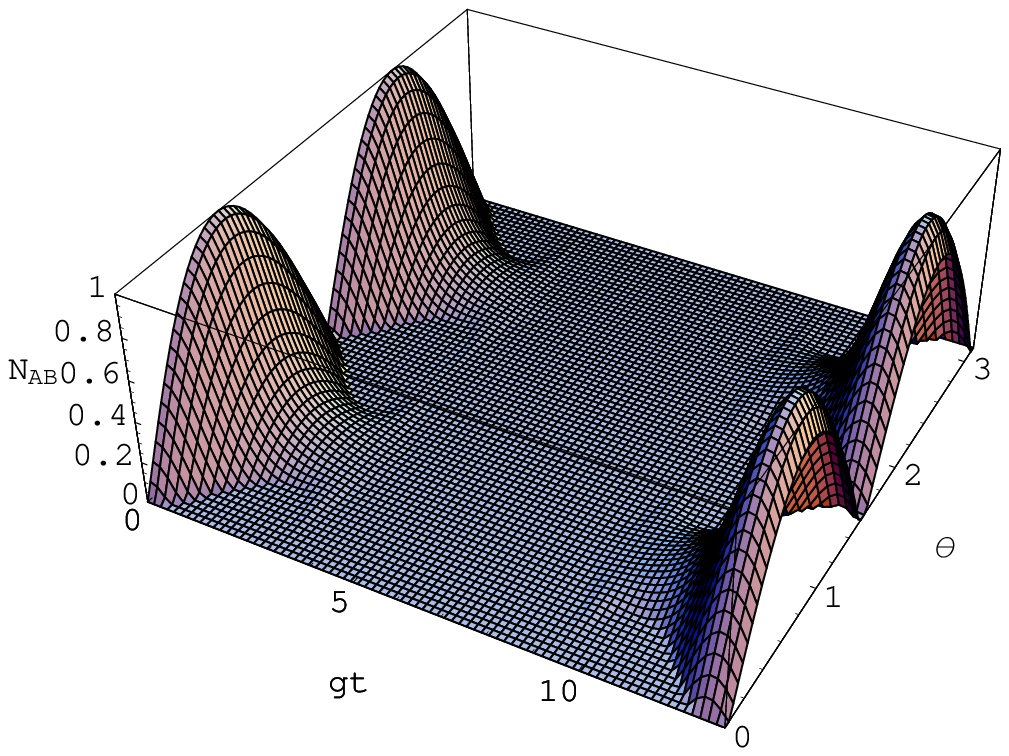}
\caption{\label{fig7} $N(\rho^{\Psi}_{AB}(t))$ as a function of
$\theta$ and the dimensionless time $gt$ for
$\delta_{A}=0.5g,\delta_{B}=2g$ in Model $2$.}
\end{figure}
\begin{figure}
\includegraphics[scale=1.2]{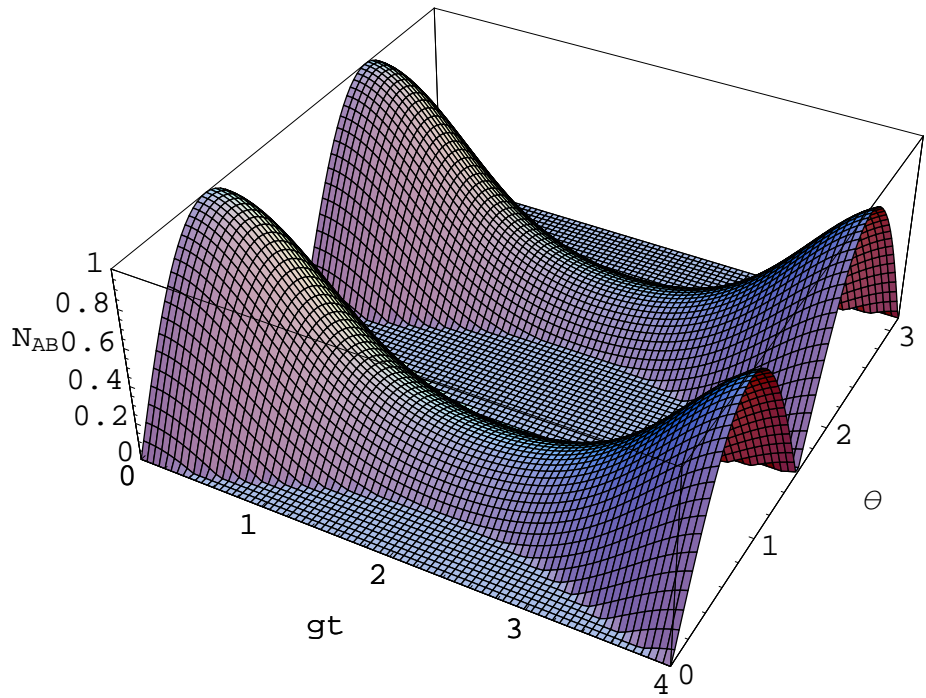}
\caption{\label{fig8} $N(\rho^{\Psi}_{AB}(t))$ as a function of
$\theta$ and the dimensionless time $gt$ for
$\delta_{A}=\delta_{B}=1.5g$ in Model $2$.}
\end{figure}
\indent The negativity of $\rho^{\Psi}_{AB}(t)$ is given by
Equation(\ref{19}) in full similarity with Model $1$, ESD always
occurs for $g_{A}/\delta_{A}>1$ and $g_{B}/\delta_{B}>1$. The
physical interpretation of the result is that $g_{A}/\delta_{A}>1$
(or $g_{B}/\delta_{B}>1$) means the strong coupling regime. Because
the atom $A$ (or $B$) and cavity $a$ (or $b$) couple with each other
strongly, the entanglement of the $AB$ subsystem can occur ESD
phenomenon. $g_{A}$ and $g_{B}$ are related to the velocity of the
initial state's disentanglement, and $\delta_{A}$, $\delta_{B}$
influence the ESD's period. In addition, the ESD also appears on the
condition $g_{A}/\delta_{A}\geq1$, $\delta_{B}{\geq}g_{B}$ or
$g_{B}/\delta_{B}\geq1$, $\delta_{A}{\geq}g_{A}$, even when
$\delta_{A}>g_{A}$, $\delta_{B}>g_{B}$ (but when $\delta_{A}>g_{A}$
or $\delta_{B}>g_{B}$, $\delta_{A}$ and $\delta_{B}$ do not
correspond to the large detuning regime), and is shown as Fig.$7$
and Fig.$8$, in which the ESD phenomenon can also occur in the case
when $\delta_{A}$ and $\delta_{B}$ are larger than $g_{A}$ and
$g_{B}$ (but not by too much). As we all know, the larger
atom-cavity detuning means that the interaction between the atom and
cavity becomes small, for example, in the large detuning regime, the
cavity and the atom can not exist energy exchange. So when
$\delta_{A}$ and $\delta_{B}$ are much larger than $g_{A}$ and
$g_{B}$, the ESD will disappear. If we consider the driven atoms and
the cavities under resonance condition $(\delta_{A}=\delta_{B}=0)$,
the negativity of $\rho^{\Psi}_{AB}(t)$ decreases at the beginning,
and then vanishes for all time (the same as Fig.$4$). So Model $2$
under the resonance regime can also be used as an information eraser.\\
 \indent For another type of atomic initial state
$|\Phi_{AB}\rangle$, it can be verified that the negativity
$N(\rho^{\Phi}_{AB}(t))$ is also equal to $N(\rho^{\Psi}_{AB}(t))$.
Through calculating the negativity $N_{Aa}$, $N_{Bb}$, $N_{Ab}$ and
$N_{Ba}$ in Model $2$, we find that the entanglement of $Aa$ and
$Bb$ can occur during the system evolution and can be on a long-time
invariable entanglement when the $AB$ subsystem is occurring the
ESD. Because of no existing interaction in the $Ba$ subsystem (or
$Ab$ subsystem), it is impossible to entangle $Ab$ (or $Ba$) in the
strongly driven regime (these results are not similar to the
standard double J-C Model $[19]$). Hence, our conclusion regarding
Model $2$ is that the driven atomic subsystem $AB$ always suffers
ESD if the atom-cavity coupling and the atom-detuning are satisfied
with the condition $g_{A}/\delta_{A}>1$, $g_{B}/\delta_{B}>1$ or
$g_{A}/\delta_{A}{\geq}1$, $\delta_{B}{\geq}g_{B}$ or
$g_{B}/\delta_{B}{\geq}1$, $\delta_{A}{\geq}g_{A}$, even when
 $\delta_{A}>g_{A}$, $\delta_{B}>g_{B}$ (but $\delta_{A}$ and
$\delta_{B}$ can not correspond to the large detuning regime),
independent of the type of the atomic initial state which may be
either $\rho^{\Phi}_{AB}(0)$ or $\rho^{\Psi}_{AB}(0)$.
\section*{$4.$ \textbf{Conclusion}}
\indent We have described the entanglement evolution of two
two-level atoms off-resonantly coupled to cavity fields. In Model
$1$, one of two atoms is trapped in a single cavity, off-resonantly
coupled to this cavity, and driven by a classic strong coherent
field, while the other remains outside the cavity and has no
environment. However, in Model $2$, each of the two strongly driven
atoms interacts with its own cavity in the absence of any coupling
between the atom-field subsystems. There are different available or
forthcoming routes to the implementation of our model. In the
microwave regime of cavity QED, pairs of atoms excited to Rydberg
levels cross a high-Q superconductive cavity with negligible
spontaneous emission during the interaction $[29]$. In the optical
regime the application of cooling and trapping techniques in cavity
QED $[23]$ allows the deterministic loading of single atoms in a
high-finesse cavity, with accurate position control and trapping
times of many seconds $[24]$. In this regime laser-assisted
three-level atoms can behave as effective two-level atoms $[30]$. On
the other hand, trapped atomic ions can remain in an optical cavity
for an indefinite time in a fixed position, where they can couple to
a single mode without coupling rate fluctuations $[31]$. These
systems are quite promising for our purposes and could become almost
ideal in the case of achievement of
the strong coupling regime.\\
\indent Under off-resonance conditions and starting from the vacuum
state of the cavity fields, for negligible atomic decays and cavity
leakage, we solved exactly the system dynamics for two types of
initial preparation of the atom pairs ($|\Phi_{AB}\rangle$ and
$|\Psi_{AB}\rangle$). Thus we found conditions for the negativity of
the so-called atomic ESD. Namely, the initial entanglement of atoms,
if any, will eventually suffer a sudden death, if the atom-cavity
coupling and the atom-cavity detuning are satisfied with
$g/\delta>1$ (in Model $1$), while in Model $2$ the system is
accordance to $g_{A}/\delta_{A}\geq1$, $g_{B}/\delta_{B}\geq1$ or
$g_{A}/\delta_{A}>1$, $\delta_{B}>g_{B}$ and/or
$g_{B}/\delta_{B}>1$, $\delta_{A}>g_{A}$ (but $\delta_{A}$ and
$\delta_{B}$ can not correspond to the large detuning regime).
Furthermore, it is interesting to note that such conditions for ESD
do not depend on the concrete type of the initial state of atoms.
The atom-cavity detuning $\delta$ influences the period of ESD and
the atom-cavity coupling $g$ is related to the
velocity of the $AB$ subsystem's disentanglement.\\
\indent In the atom-cavity resonance regime, the negativity of the
atomic subsystem decreases at the beginning, and then vanishes for
any time, so the strongly driven model (such as Model $1$ and Model
$2$) under the resonance regime can be used as an information
eraser. Entangled states are generally very fragile against
interaction with environments, but in our present scheme, the energy
exchange between atoms and cavities do not exist under the
atom-cavity large detuning regime, so the initial entanglement of
atoms is preserved during system evolution. That is to say, the
strongly driven and large detuning model can be used to preserve the
information (entanglement) in
quantum information processing and quantum computing.\\
\indent In the strongly driven regime, we note the negativity of two
atoms is a periodic function of time in the close system. It means
that atoms evolve quasiperiodically between the entangled and
disentangled states. This result is different from refs.$[15,16]$,
ref.$[15]$ shows that the negativity of the bipartite system is a
mononically decreasing function of time when the first qubit is
trapped in a infinite heat bath and the second qubit has no
environment. But authors in ref.$[15]$ also have proposed that in
the case when the second qubit is set in a finite and controlling
quantum environment, the dynamics of entanglement can oscillate,
this nonmontonic behavior appears to be linked to the non-markovian
character of the dynamics. That is to say, non-markovian
entanglement can evolve quasiperiodically between the entangled and
disentangled states in the open systems. From the paper $[15]$ we
acquire the degree of entanglement at certain time intervals can be
manipulated by means of an appropriate choice of the initial state
of the controlling quantum environment. Our results in the strongly
driven regime manifest the entanglement of two atoms can be
manipulated through controlling the atom-cavity detuning and
atom-cavity coupling.
\section*{\textbf{Acknowledgments}}
YJZ, ZXM and YJX are supported by the National Science Foundation of
China under Grant No.10774088, the Key Program of National
 Science Foundation of China under Grant No.10534030.
 \end{document}